\documentclass[journal=jctcce,manuscript=article,email=false]{achemso}

\usepackage{amsmath,amsfonts,amssymb}
\usepackage{array, booktabs, tabularx}
\usepackage{xcolor, graphicx, tikz}
\usepackage[caption=false]{subfig}
\usepackage{hyperref} 
\usepackage{braket}
\usepackage{bm}

\usepackage{makecell}
\usepackage{booktabs}

\usepackage{qcircuit}

\usepackage{algpseudocode}
\usepackage{algorithm}
\usepackage{siunitx}
\usepackage{caption}
\usepackage[export]{adjustbox}

\newcommand*{\vv}[1]{\vec{\mkern0mu#1}}

\title{Subspace-Search Quantum Imaginary Time Evolution for Excited State Computations}

\author{Cameron Cianci}
\affiliation{Physics Department, University of Connecticut, Storrs, Connecticut 06269, USA}
\alsoaffiliation{Mirion Technologies (Canberra) Inc., 800 Research Parkway, Meriden, Connecticut 06450, USA}
\email{cameron.cianci@uconn.edu}
\author{Lea F. Santos}
\affiliation{Physics Department, University of Connecticut, Storrs, Connecticut 06269, USA}
\author{Victor S. Batista}
\affiliation{Department of Chemistry, Yale University, P.O. Box 208107, New Haven, Connecticut 06520-8107, USA}
\alsoaffiliation{Yale Quantum Institute, Yale University, New Haven, Connecticut 06511, USA}

\begin{document}

\begin{abstract}
Quantum systems in excited states are attracting significant interest with the advent of noisy intermediate scale quantum (NISQ) devices. While ground states of small molecular systems are typically explored using hybrid variational algorithms like the variational quantum eigensolver (VQE), the study of excited states has received much less attention, partly due to the absence of efficient algorithms. In this work, we introduce the {\em subspace search quantum imaginary time evolution} (SSQITE) method, which calculates excited states using quantum devices by integrating key elements of the subspace search variational quantum eigensolver (SSVQE) and the variational quantum imaginary time evolution (VarQITE) method. The effectiveness of SSQITE is demonstrated through calculations of low-lying excited states of benchmark model systems, including $\text{H}_2$ and $\text{LiH}$ molecules. A toy Hamiltonian is also employed to demonstrate that the robustness of VarQITE in avoiding local minima extends to its use in excited state algorithms. With this robustness in avoiding local minima, SSQITE shows promise for advancing quantum computations of excited states across a wide range of applications.
\end{abstract}

\section{Introduction}

Computational and theoretical studies of excited states are essential for understanding the photophysics of molecules, particularly in UV-vis and X-ray absorption spectroscopy of photochemical reactions.\cite{olivucci2005computational,josefsson2012ab} With the advent of quantum computing, new methodologies promise to significantly enhance these studies, potentially offering a quantum advantage in chemistry.~\cite{bauer2020quantum,motta2022emerging} Traditional computational methods, despite their powerful capabilities, face limitations in modeling complex excited-state phenomena due to the exponential scaling of resources required. Quantum computing, however, opens new frontiers for exploring a wide range of problems~\cite{Feynman1982,RevModPhys86153}, including the crucial excited states in the photochemistry of organic molecules.~\cite{Klessinger1995ExcitedSA}  

In the near-term intermediate scale quantum (NISQ) era, quantum advantage of some specialized applications have already been put forward~\cite{Daley2022, Kim2023}, such as the calculation of ground-state energy in quantum chemistry~\cite{Tilly_2022,lee2023evaluating, Peruzzo_2014}. Widespread approaches for calculating ground-state energies in quantum computers include the hybrid variational quantum eigensolver (VQE) algorithm~\cite{Tilly_2022, Peruzzo_2014, McArdle_2019}, or the variational quantum imaginary time evolution (VarQITE) method~\cite{McArdle_2019,Kyaw_2023, PhysRevResearch.3.033083}. Beyond ground-state energies, 
excited states are equally important for numerous applications~\cite{ExcitedStateProgress, SERRANOANDRES200599, Matsika2018, Benavides_Riveros_2024, doi:10.1021/acs.jctc.2c00446, PhysRevA.108.022814}, such as charge and energy transfer in photovoltaic materials, photodissociation \cite{Higgott_2019}, luminescence \cite{Klessinger1995ExcitedSA}, intermediate states in chemical reactions \cite{Lischka2018}, and mechanistic studies of catalytic systems \cite{Reiher_2017}. This has driven significant interest in generalizing ground state algorithms such as VQE and VarQITE to excited states of quantum systems. Notable algorithms designed for this purpose include the subspace-search variational quantum eigensolver (SSVQE) \cite{Nakanishi_2019} and the variational quantum deflation (VQD) \cite{Higgott_2019} algorithm. The VQD approach~\cite{Higgott_2019} has been applied to calculations at Frank–Condon and the conical intersection geometries,~\cite{gocho2023excited} and has been adapted to VarQITE \cite{Jones_2019, Tsuchimochi_2023} for determining excited states.  

Quantum algorithms for imaginary time evolution have proven useful in the determination of both ground and excited states.  There are two quantum algorithms which can perform imaginary time evolution in quantum computers, variational quantum imaginary time evolution (VarQITE) \cite{McArdle_2019, doi:10.1021/acsomega.3c01060, PhysRevResearch.4.033121}, and trotterized quantum imaginary time evolution (TrotterQITE) \cite{Motta_2019}.  VarQITE uses a variational circuit to approximate the evolution of the input state through imaginary time, whereas TrotterQITE implements a non-unitary imaginary time step $e^{-\mathcal{H}d\tau}$ by applying a normalized unitary timestep $e^{-iAd\tau}$ with ancilla qubits.  Due to the ancilla qubits, TrotterQITE requires many more qubits and a larger gate depth than VarQITE. Due to the fixed gate depth and therefore greater noise resilience of VarQITE, this algorithm is used in the excited state algorithm presented.

Imaginary time algorithms have been applied to determine excited states through other methods, such as subspace expansion methods\cite{motta2023subspacemethodselectronicstructure}.  Subspace expansion methods define a subspace of the system using non-orthogonal states, with the exception of multistate contracted VQE\cite{Parrish_2019}, and classically diagonalize this subspace using the generalized eigenvalue equation \cite{stair2019multireferencequantumkrylovalgorithm, PhysRevX8011021, parrish2019quantumfilterdiagonalizationquantum, Parrish_2019}, rather than minimizing the entire Lagrangian as is performed in VQE or VarQITE.  These subspace expansion methods perform well when the input states have large overlap with the low-energy states of interest.  For this reason, TrotterQITE has been used in conjunction with subspace expansion, referred to as Krylov subspace methods \cite{Motta_2019, motta2023subspacemethodselectronicstructure, tkachenko2023quantumdavidsonalgorithmexcited, PhysRevA.105.022417}.  These algorithms have accuracy guarantees, but can require deep quantum circuits to perform TrotterQITE.  Comparing subspace expansion to subspace search, the subspace expansion does not yield an orthonormal set of states, whereas subspace search ensures orthogonality of the output states.

In this paper, we introduce a novel algorithm called subspace search quantum imaginary time evolution (SSQITE). The SSQITE algorithm augments VarQITE with subspace search to computing excited states to enable the simultaneous calculation of ground and multiple excited states. Its efficiency is successfully demonstrated with the calculation of the low-lying states of $\text{H}_2$ and $\text{LiH}$ molecules. The paper is organized as follows. First, we introduce the SSVQE and VarQITE methods in Secs.~\ref{sec:SSVQE} and~\ref{sec:QITE}, respectively. Then, we describe the SSQITE algorithm in Sec.~\ref{sec:SSQITE} and illustrate its application to calculations of excited states of H$_2$ and $\text{LiH}$, as well as introduce a toy Hamiltonian to demonstrate SSQITE's robustness to local minima in Sec.~\ref{sec:h2lih}. Conclusions are presented in Sec.~\ref{sec:conclusions}.

\section{Subspace-Search Variational Quantum Eigensolver}
\label{sec:SSVQE}

The subspace-search variational quantum eigensolver (SSVQE) algorithm extends the variational quantum eigensolver (VQE) hybrid method \cite{Peruzzo_2014, Tilly_2022}. The VQE is a hybrid quantum-classical algorithm designed to find the ground state of a quantum system described by the $2^n \times 2^n$ Hamiltonian, $H$, expressed as a sum of tensor products of Pauli matrices $\sigma_k^{(j)}=\{X, Y, Z, I\}$, 
\begin{equation}
    H=\sum_j c_j \bigotimes_{k=1}^n \sigma_k^{(j)},
\label{eq:ham}
\end{equation}
where $c_j=2^{-n}~\text{Tr}[H \times \bigotimes_{k=1}^n \sigma_k^{(j)}]$. VQE generates a trial state $\vert \psi(\vv{\theta}) \rangle = U(\vv{\theta}) \vert \psi_0 \rangle$ by applying a quantum circuit $U(\vv{\theta})$ with variational parameters $\vv{\theta}$ to an initial vacuum state $\vert \psi_0 \rangle$. These parameters are adjusted by a classical computer to minimize the expectation value of the Hamiltonian, $E(\vv{\theta})=\langle \psi(\vv{\theta})| \hat{H} \vert \psi(\vv{\theta}) \rangle$. This expectation value is computed by summing the expectation values of the tensor products of Pauli matrices, $\langle \psi(\vv{\theta}) |\bigotimes_{k=1}^n \sigma_k^{(j)} \vert \psi(\vv{\theta}) \rangle$, measured on the quantum computer. The process iteratively refines $\vv{\theta}$ to minimize $E(\vv{\theta})$, thereby approximating the lowest eigenvalue of $H$.

SSVQE extends the VQE algorithm to simultaneously find the $k$ lowest eigenstates of $H$~\cite{Nakanishi_2019}. First, $k$ orthogonal states $\ket{\phi_j}$ are initialized with $\langle \phi_k\ket{\phi_j}=\delta_{kj}$. These states are then evolved using the same circuit $U(\vv{\theta})$ with variational parameters $\vv{\theta}$. Orthogonality is thus preserved among the evolved states since $U(\vv{\theta})^\dagger U(\vv{\theta})=I$, so $\bra{\phi_k} U(\vv{\theta})^\dagger U(\vv{\theta}) \ket{\phi_j}=\delta_{kj}$. The ansatz defining the circuit $U(\vv{\theta})$ can be chosen to preserve the symmetry, such as the “ASWAP” ansatz which is constructed using gates that preserve the number of excitations in a state.~\cite{barron2021preserving}

The parameters $\vv{\theta}$ are optimized by minimizing the sum of the expectation values using the following loss function:
\begin{equation}
\label{eq:loss}
    \mathcal{L}_\omega (\vv{\theta}) = \sum_{j=0}^k \omega_j \bra{\phi_j}U^{\dag}(\vv{\theta})HU(\vv{\theta})\ket{\phi_j}.
\end{equation}
Therefore, SSVQE finds the $k$ orthogonal minimum energy states simultaneously. The coefficients $\omega_i$, introduced by Eq.~(\ref{eq:loss}), with $\omega_i > \omega_j$ for $i < j$, are used to weight each energy level, effectively arranging the energy expectation values of all orthogonal states in ascending order.  

In this paper, we introduce the subspace search quantum imaginary time evolution (SSQITE) algorithm by integrating this SSVQE methodology of orthogonal states with the VarQITE algorithm.~\cite{McArdle_2019,Kyaw_2023} The resulting SSQITE method thus enables the simultaneous calculation of multiple excited states by applying the same imaginary time evolution to an initial set of orthogonal states.

\section{Variational Quantum Imaginary Time Evolution}
\label{sec:QITE}

The variational quantum imaginary time evolution (VarQITE) algorithm is a hybrid quantum-classical method used to determine the ground-state energy of a quantum system by propagating an initial state $\ket{\psi(0)}$ in imaginary time toward $\ket{\psi(\tau)}$, where $\tau = i t/\hbar$ is the imaginary time~\cite{McArdle_2019,Kyaw_2023}. This technique effectively implements the Wick-rotated Schr\"{o}dinger equation,
\begin{equation}
    \frac{d}{d\tau}\ket{\psi(\tau)} = -(\mathcal{H} - E_{\tau})\ket{\psi(\tau)},
\end{equation}
with $E_{\tau} = \bra{\psi(\tau)}\mathcal{H}\ket{\psi(\tau)}$.  Propagating that initial state for a sufficiently long imaginary time, we obtain the ground-state $\ket{E_0}$, provided that $\langle E_0\ket{\psi(0)} \neq 0$. This is expressed, as follows:
\begin{equation}
    \lim_{\tau \to \infty} A(\tau) e^{- H \tau} \ket{\psi(0)} = \ket{E_0},
\end{equation}
where $A(\tau)= \bra{\psi(0)} e^{- 2 H \tau} \ket{\psi(0)}^{-1/2}$ is the normalization factor obtained after imaginary-time propagation. To apply this procedure to a given parameterized ansatz $\ket{\psi(\tau)}=U(\theta(\tau))\ket{0}$, McLachlan's variational principle can be leveraged, which states:
\begin{equation}
\label{eq:mcl}
    \delta \left|\left|\left(\frac{d}{d\tau} + \mathcal{H} - E_{\tau}\right)\ket{\psi(\tau)}\right|\right| = 0.
\end{equation}
Applying this principle to the optimization of the variational parameters $\vv{\theta}$ that define $U(\vv{\theta}(\tau))$ results in the following linear system of ordinary differential equations~\cite{McArdle_2019,Kyaw_2023}: 
\begin{equation}
    \label{eq:euler}
    \sum_j A_{ij} \dot\theta_j = C_i,
\end{equation}
where
\begin{equation}
    A_{ij} = \Re\left(\frac{\partial\bra{\phi(\vv{\theta}(\tau))}}{\partial\theta_i}\frac{\partial\ket{\phi(\vv{\theta}(\tau))}}{\partial\theta_j}\right),
\end{equation}
and
\begin{equation}
    C_i = - \Re\left(\bra{\frac{\partial\phi(\vv{\theta}(\tau))}{\partial\theta_i}}\mathcal{H}\ket{\phi(\vv{\theta}(\tau))}\right).
\end{equation}
The values of $A_{ij}$ and $C_i$ are obtained using the Hadamard test on a quantum circuit by simply averaging the measurements on the ancilla qubit ~\cite{McArdle_2019}.  

Having obtained $A_{ij}$ and $C_i$ by measurements of the ancilla in the quantum circuit, the values of $\vv{\theta}$ are updated in a classical computer by integrating the Euler equation introduced by Eq.~(\ref{eq:euler}) using the 4th-order Runge-Kutta method~\cite{runge1895numerische}. The process is iterated until the values of $\vv{\theta}$ converge to optimum values, as determined by the McLachlan's variational principle introduced by Eq.~\ref{eq:mcl}.

\section{Subspace-Search Quantum Imaginary Time Evolution}
\label{sec:SSQITE}

The subspace-search quantum imaginary time evolution (SSQITE) method, proposed in this paper, combines subspace search optimization with variational quantum imaginary time evolution to maintain orthogonality among states evolving in imaginary time. This approach allows for the simultaneous variational computation of both ground and excited energy states using variational quantum imaginary time evolution.

The main difficulty in combining the subspace search optimization with variational quantum imaginary time evolution is that the imaginary time propagation only implicitly optimizes the loss function defined by the McLachlan's variational principle in Eq.~(\ref{eq:mcl}).  Instead of defining a joint loss function, as in SSVQE, the SSQITE algorithm tunes the step size $d\tau_j$ of each level $j$ individually, such that lower energy states have larger integration time steps (pseudo-code, Algorithm 1).  Intuitively, this allows for lower energy states to overpower the higher energy states, ordering the output energy spectrum.  The tuning of time-steps plays a similar role as the tuning of the weights $\omega_i$ in the SSVQE algorithm.  In this way, after a sufficient number of iterations, the SSQITE algorithm returns the $k$-lowest-energy eigenstates.

The choice of weights $\omega_i$ can greatly impact the convergence of the algorithm, and has been previously been chosen to take advantage of the choice of input states and ansatz, system size, or symmetries \cite{Benavides_Riveros_2022, Benavides_Riveros_2024}.  For example, the weight selection for fastest convergence of CQE on H$_2$ was found to be $\omega_i = [9,9,1,1]$ \cite{Benavides_Riveros_2024}, as it takes advantage of the block-diagonal nature of the Hamiltonian.  Here we will instead demonstrate a weight setting scheme which utilizes the nature of orthogonal states evolving under VarQITE to prevent the evolution of higher energy states from overpowering lower energy states, while retaining an efficient runtime.

\begin{algorithm}
\caption{Pseudo-code for the SSQITE Algorithm}
\label{alg:cap}
\begin{algorithmic}
\Require $\psi = \psi_i,~\text{with}~0 \leq i < k.$
\Ensure $\langle\psi_i\ket{\psi_j} = \delta_{ij}$
\While{not all\_converged(${\dot\theta}$)}
    \State $d\tau_i \gets \{\frac{1}{2^i} | 0 \leq i < k\}$ \Comment{Initialize Step Sizes}
    \State $A_{ijl} \gets$ Measure\_A($U(\vv{\theta})\psi_l$) 
    \State $C_{il} \gets$ Measure\_C($U(\vv{\theta})\psi_l$)
    \State $\dot\theta_{jl} \gets A^{-1}_{ijl}C_{il}$ \Comment{Calculate $\dot\theta$}
    \For{$l = 0$, $l < k$, $l++$}
        \If{converged($\dot\theta_l$)}
            \For{$i=l$, $i<k$, $i++$}
                \State $d\tau_i \gets 2*d\tau_i$ \Comment{Avoid Exponential Scaling with $k$}
            \EndFor
        \EndIf
    \EndFor
    \For{$j = 0$, $j<num\_params$, $j++$}
        \For{$l = 0$, $l<k$, $l++$}
            \State $\theta_{jl} \gets \theta_{jl} + d\tau_{l}*\dot\theta_{jl}$ \Comment{Update Theta}
        \EndFor
    \EndFor
\EndWhile
\end{algorithmic}
\end{algorithm}


In this weight setting scheme, the integration time steps are defined as follows:
\begin{equation}
    d\tau_i = \frac{b}{2^i}.
\end{equation}
with $b$ a tunable parameter. This choice of integration time steps prevents higher energy levels from overpowering lower energy eigenstates, since 
\begin{equation}
    \frac{1}{2^i} \geq \sum_{j=i+1}^k \frac{1}{2^j}.
\end{equation}
However, this approach requires a number of steps that scales exponentially as $\mathcal{O}(2^{k})$, where $k$ is the size of the subspace.  This exponential scaling can be overcome by leveraging the convergence of lower energy levels. The integration time steps used for obtaining higher energy levels can be increased upon convergence of lower energy states since all remaining states must be orthogonal to the manifold of lower energy states $\langle E_j\ket{\psi_i} \approx \delta_{ji}$ for $i>j$.  Therefore, the imaginary time evolution of higher excited states is restricted to an orthogonal subspace.

Due to the time evolution of excited states being restricted, the integration time step of these states can be doubled, mitigating the exponential scaling without significantly affecting the lower energy states.  However, the imaginary time evolution of the ground state makes the overlap with excited states exponentially small, although not exactly zero $\langle E_0 \ket{\psi_i} \approx e^{-\tau}$.  Therefore, in practice, some excited states can still evolve into the ground state if they are not fully orthogonalized. So, it is always necessary to confirm orthogonality with lower energy states during each round of SSQITE. 

\section{Results: Ground and Excited states of $\text{H}_2$ and $\text{LiH}$}
\label{sec:h2lih}

\begin{figure}[t]
\begin{center}
    \includegraphics[scale = 0.6]{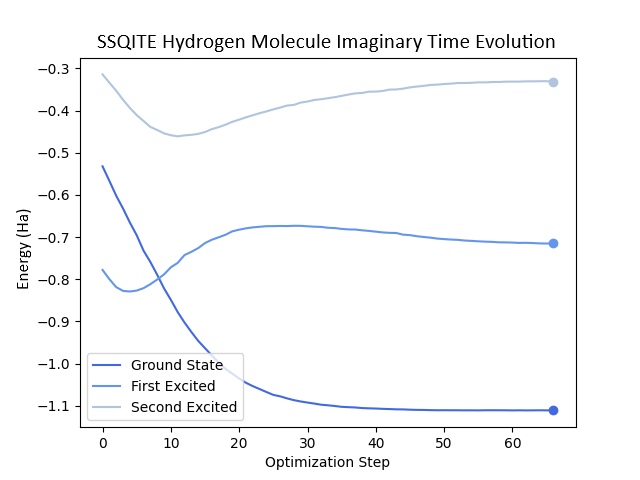}
\end{center}
    \caption{Simultaneous evolution of the energy expectation values for the three lower energy states of H$_2$ (with fixed bond-length $R=0.95 \AA$) during the first 70 integration steps of SSQITE optimization. Final energy values are highlighted on the right, and corresponding statistical errors are of the order $10^{-5}$ Ha.}
\label{fig:fig1o}
\end{figure}

SSQITE was implemented on H$_2$ using a two-qubit Hamiltonian. This H$_2$ Hamiltonian was created by beginning with the STO-3G basis, and selecting for the spin zero subspace to provide four states, which can be directly mapped to two-qubit states \cite{PhysRevX8011021}. This Hamiltonian has previously been used in conjunction with quantum subspace expansion, achieving an error far exceeding chemical accuracy for a range of interatomic distances\cite{PhysRevX8011021}.

Figure~\ref{fig:fig1o} illustrates the energy expectation values for the three lowest energy states of H$_2$ during joint SSQITE optimization (with a fixed H-H bond-length of $0.95 \AA$). The imaginary time propagation causes these states to interfere through their contributions to $\vv{\dot{\theta}}$.  As shown in Fig.~\ref{fig:fig1o}, the evolution of the ground state for $\tau \in [0,20]$ leads to an increase in the energy of the first excited state, as it is forced into a subspace orthogonal to the ground state.  This effect is also reciprocal, since the evolution of the first excited state likely slows the evolution of the ground state, as evidenced by the linear slope of the ground state from $\tau=0$ to $\tau=15$.

\begin{figure}[t]
(a)
\includegraphics[scale=0.45]{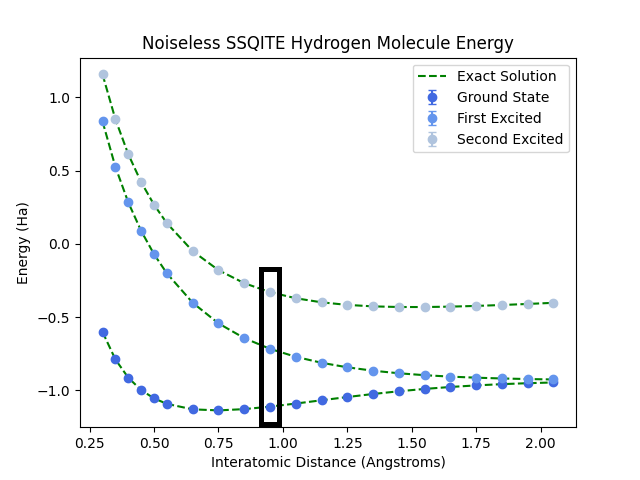} 
(b)
\includegraphics[scale=0.45]{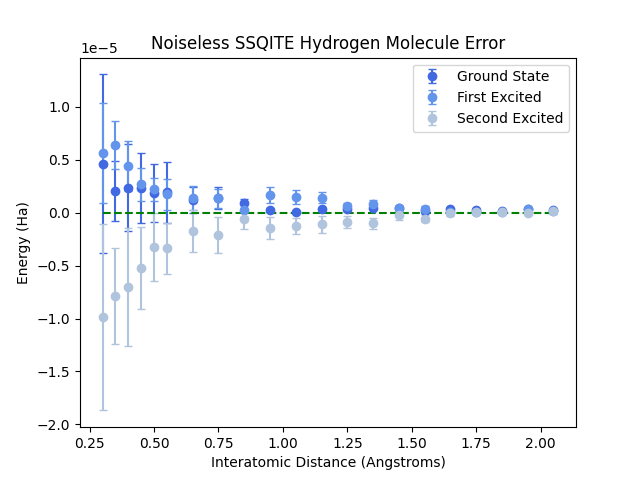}
(c)
\includegraphics[scale=0.45]{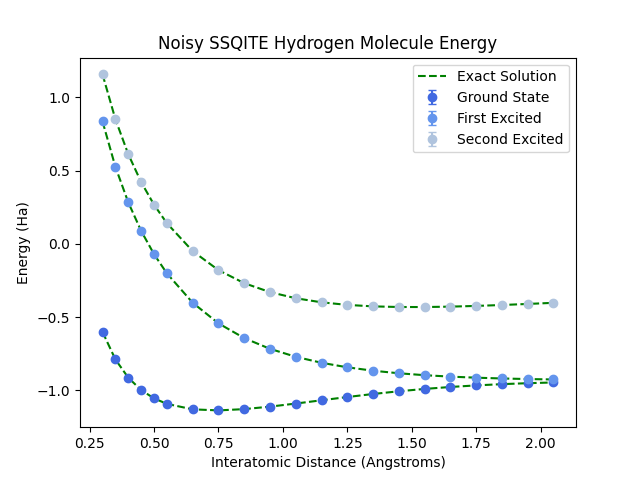} 
(d)
\includegraphics[scale=0.45]{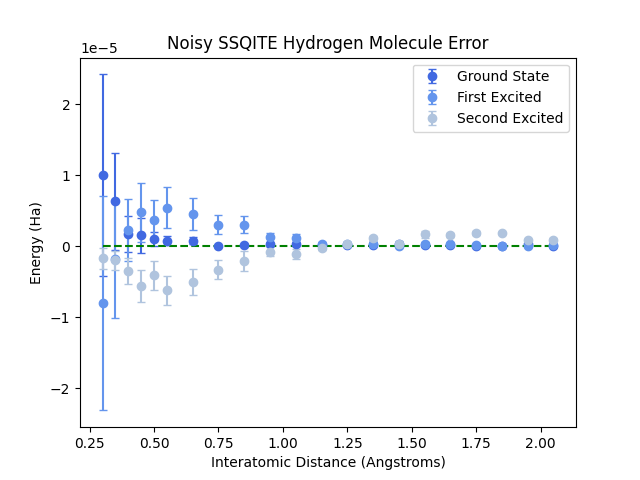}
\caption{Comparison of the three lowest energy eigenvalues of $\text{H}_2$ determined through (a)-(b) noiseless and (c)-(d) noisy SSQITE optimization to numerically exact calculations (dashed lines) as a function of the interatomic $\text{HH}$ distance. Boxed values correspond to the final values shown highlighted in Fig.~\ref{fig:fig1o}.  The ground, first, and second excited states correspond to the X$^1\Sigma_g^+$, b$^3\Sigma_u^+$, and B$^1\Sigma_u^+$ states of H$_2$ respectively.  Deviations of (b) noiseless and (d) noisy SSQITE calculations from the ground truth energy levels of the $\text{H}_2$ molecule.  All noisy simulations are performed using the qiskit FakeSherbrooke backend.}
\label{fig:fig2o}
\end{figure}
%
\begin{figure}[h!]
\Qcircuit @C=0.5em @R=1.5em {
 \lstick{\ket{0}} & \gate{R_X(\theta_1)} & \gate{R_Y(\theta_3)} & \ctrl{1} & \gate{R_X(\theta_5)} & \gate{R_Y(\theta_7)} & \ctrl{1} & \gate{R_X(\theta_9)} & \gate{R_Y(\theta_{11})} & \ctrl{1} & \gate{R_X(\theta_{13})} & \gate{R_Y(\theta_{15})} & \qw\\
 \lstick{\ket{0}} & \gate{R_X(\theta_2)} & \gate{R_Y(\theta_4)} & \targ    & \gate{R_X(\theta_6)} & \gate{R_Y(\theta_8)} & \targ    & \gate{R_X(\theta_{10})} & \gate{R_Y(\theta_{12})} & \targ    & \gate{R_X(\theta_{14})} & \gate{R_Y(\theta_{16})} & \qw\\
}
\caption{Variational quantum circuit ansatz with two qubits used for the SSQITE H$_2$ calculations shown in Fig.~\ref{fig:fig2o}. The TwoLocal ansatz involves one layer of parameterized \text{RX} and \text{RY} gates, followed by a \text{CNOT} gate.  This ansatz is general, in the sense that it can realize any two-qubit operation.}
    \label{fig:fig22o}
\end{figure}
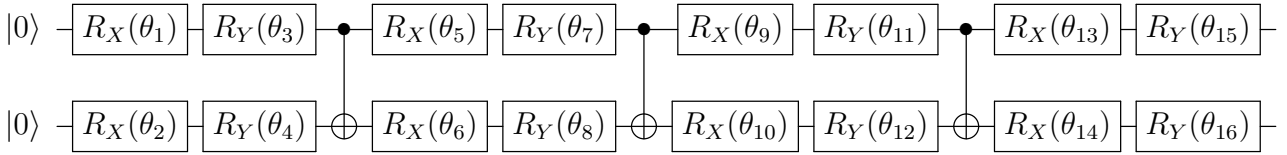

Figures~\ref{fig:fig2o}(a),(c) show the three lowest energy eigenvalues of $\text{H}_2$ determined through SSQITE optimization.  These calculations use a general two-qubit ansatz depicted in Figure~\ref{fig:fig22o}, as a function of the interatomic $\text{H-H}$ distance \cite{PhysRevX8011021}. These results demonstrate excellent agreement with exact results on both noiseless (Figure~\ref{fig:fig2o}(a)) and noisy (Figure~\ref{fig:fig2o}(c)) quantum simulators. In fact, the comparison to numerically exact calculations shown in Figures~\ref{fig:fig2o}(a),(c) demonstrates the accuracy and capabilities of the SSQITE algorithm over the entire range of bond-lengths.

Figure~\ref{fig:fig2o}(b) [Figure~\ref{fig:fig2o}(d)] shows the errors of the noiseless [noisy] SSQITE calculations for the H$_2$ molecule, which remain within \SI{9.8e-6}{Ha} (\SI{1.0e-5}{Ha}), i.e., within chemical accuracy of \SI{1.6e-3}{Ha}~\cite{Higgott_2019}.

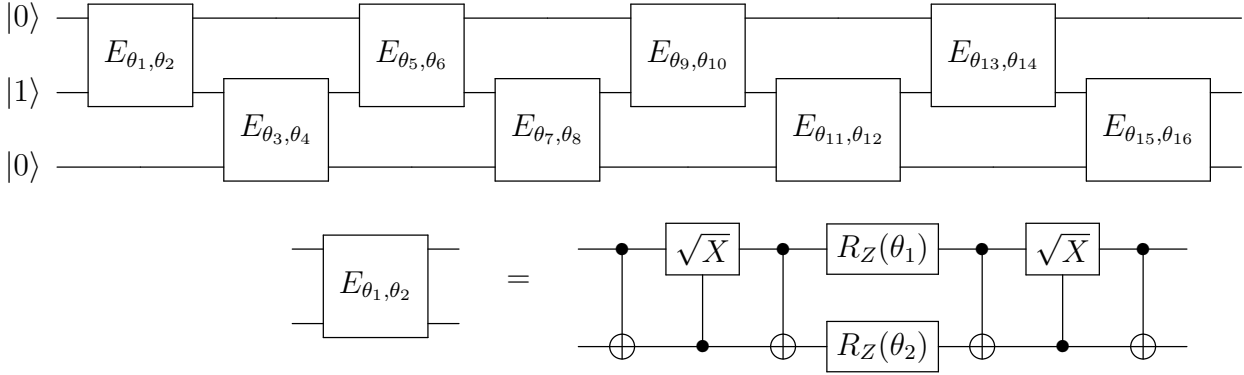
\begin{figure}[h]
\Qcircuit @C=1em @R=1.5em {
\lstick{\ket{0}} & \multigate{1}{E_{\theta_1, \theta_2}} &  \qw              & \multigate{1}{E_{\theta_5, \theta_6}} & \qw             & \multigate{1}{E_{\theta_9, \theta_{10}}} & \qw              & \multigate{1}{E_{\theta_{13}, \theta_{14}}} & \qw              & \qw \\
\lstick{\ket{1}} & \ghost{E_{\theta_1, \theta_2}}        & \multigate{1}{E_{\theta_3, \theta_4}} & \ghost{E_{\theta_5, \theta_6}}        & \multigate{1}{E_{\theta_7, \theta_8}} & \ghost{E_{\theta_9, \theta_{10}}}        & \multigate{1}{E_{\theta_{11}, \theta_{12}}} & \ghost{E_{\theta_{13}, \theta_{14}}}        & \multigate{1}{E_{\theta_{15}, \theta_{16}}} & \qw \\
\lstick{\ket{0}} & \qw              & \ghost{E_{\theta_3, \theta_4}}        & \qw              & \ghost{E_{\theta_7, \theta_8}}        & \qw              & \ghost{E_{\theta_{11}, \theta_{12}}}        & \qw              & \ghost{E_{\theta_{15}, \theta_{16}}}        & \qw  \\
}
\begin{tabular}{ccc}
\hspace{1.5cm}
\Qcircuit @C=1em @R=1.5em {
 & \multigate{1}{E_{\theta_1, \theta_2}} & \qw \\
 & \ghost{E_{\theta_1, \theta_2}}        & \qw \\
} &
\begin{tabular}{c}
 \\
 \\
=\\
\end{tabular}  &
\Qcircuit @C=1em @R=1.5em {
 & \ctrl{1} & \gate{\sqrt{X}} & \ctrl{1} & \gate{R_Z(\theta_1)} & \ctrl{1} & \gate{\sqrt{X}} & \ctrl{1} & \qw\\
 & \targ    & \ctrl{-1}       & \targ    & \gate{R_Z(\theta_2)} & \targ    & \ctrl{-1}       & \targ    & \qw\\
}\\
\end{tabular}
\caption{Top: Variational quantum circuit ansatz with three qubits used for the SSQITE $\text{LiH}$ calculations shown in Fig.~\ref{fig:fig4} based on a custom excitation preserving ansatz. Bottom: Excitation preserving subcircuit with two tunable parameters.}
\label{fig:fig7}
\end{figure}

For comparison, we also apply the SSQITE algorithm to the $\text{LiH}$ molecule \cite{Zong_2024}, using a custom excitation preserving ansatz with 16 adjustable parameters shown in Figure~\ref{fig:fig7}.  This excitation preserving ansatz ensures that the occupation number symmetry is preserved by SSQITE. The three-qubit $\text{LiH}$ Hamiltonian is obtained by beginning with the STO-6G basis. Reducing the size of the active space down to three orbitals based on the natural orbital occupation number (NOON) and averaging the qubits, we are left with a three-qubit $\text{LiH}$ Hamiltonian under the STO-6G basis.~\cite{Zong_2024,PhysRevX.8.031022} We note that the model Hamiltonian studied here involves a representation of the $\text{LiH}$ based on a truncated atomic orbital basis set that includes only $s$ orbitals.~\cite{Cadi_Tazi_2024}
To match the experimental values for the $\text{LiH}$ molecule, extended basis sets need to be incorporated into its Hartree--Fock calculations,~\cite{kahalas1963electronic,tung2011very} which is outside the scope of this paper.

%
\begin{figure}[h!]
(a)
\includegraphics[scale=0.45]{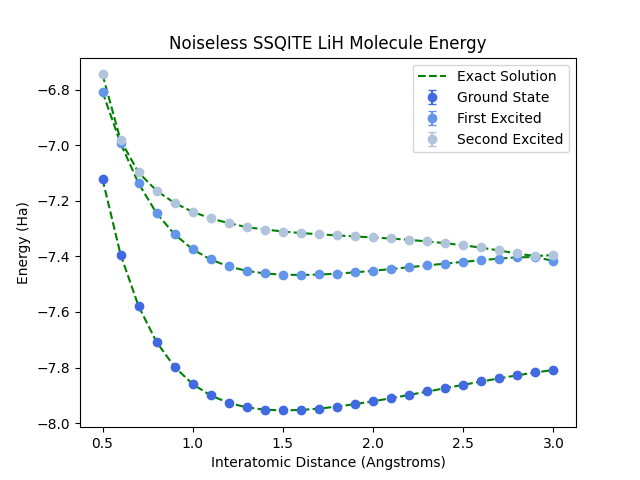}
(b)
\includegraphics[scale=0.45]{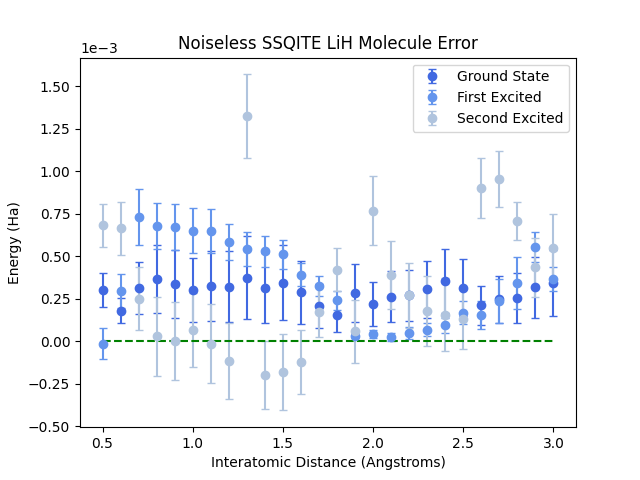}
(c)
\includegraphics[scale=0.45]{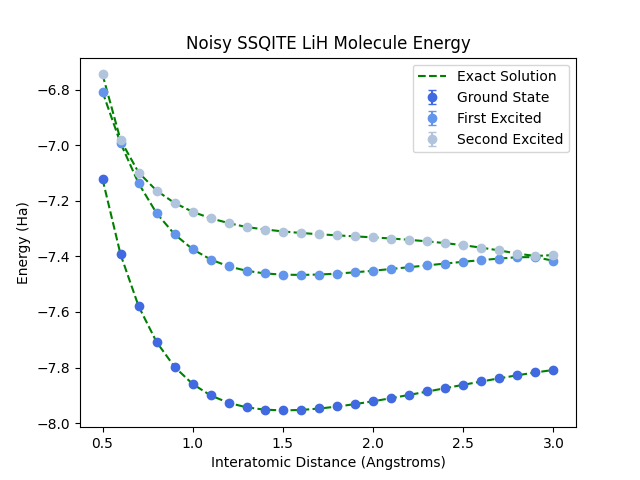} 
(d)
\includegraphics[scale=0.45]{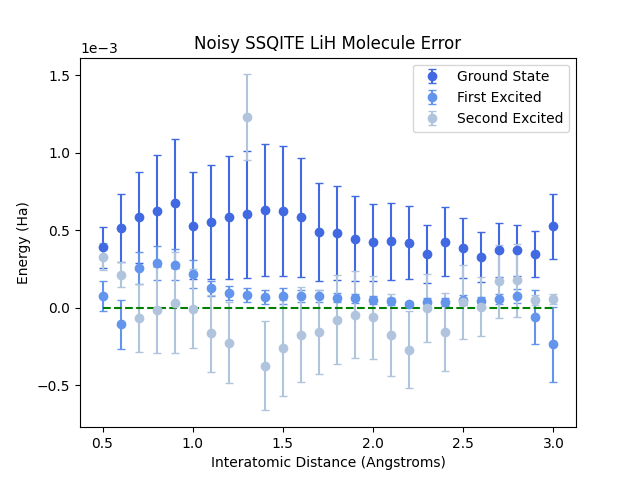}
\caption{Comparison of the three lowest energy eigenvalues of $\text{LiH}$ determined through (a)-(b) noiseless and (c)-(d) noisy simulation of SSQITE optimization to numerically exact calculations (dashed lines). The ground, first, and second excited states correspond to X$^1\Sigma^+$, a$^3\Sigma^+$, and A$^1\Sigma^+$ respectively. Note that the results from LiH differ from experimental data due to the truncated atomic orbital basis set used. Depicted are the Deviations of the (b) noiseless and (d) noisy SSQITE calculations from the ground truth energy levels of the $\text{LiH}$ molecule.  All noisy simulations are performed using the qiskit FakeSherbrooke backend.}
\label{fig:fig4}
\end{figure}

Figures~\ref{fig:fig4}(a),(c) show the three lowest energy eigenvalues of $\text{LiH}$ as a function of the interatomic $\text{Li-H}$ distance for the noiseless [Figure~\ref{fig:fig4}(a)] and noisy [Figure~\ref{fig:fig4}(c)] SSQITE optimization. The results show excellent agreement with benchmark calculations of the entire range of interatomic distances.

Figures~\ref{fig:fig4}(b),(d) shows the errors of SSQITE calculations for the $\text{LiH}$ model, which remain within chemical accuracy. Similarly to the performance for the $\text{H}_2$ molecule, SSQITE performs well in calculations of ground and excited state energies of $\text{LiH}$. In fact, as shown in Fig.~\ref{fig:fig4}, the noiseless (noisy) algorithm exhibits a maximum deviation of \SI{1.30e-3}{Ha} (\SI{1.32e-3}{Ha}), below the benchmark of \SI{1.6e-3}{Ha}. The noisy results perform remarkably similar to the noiseless results for both H$_2$ and $\text{LiH}$ for two reasons.  Firstly the circuits employed have limited gate depth and are therefore resistant to noise.  Secondly, the added gate noise delays convergence of the algorithm, reducing the effect of the noise at the cost of a small number of extra iterations.

\begin{figure}[h!]
(a)\hspace{1cm}
\includegraphics[scale=0.25]{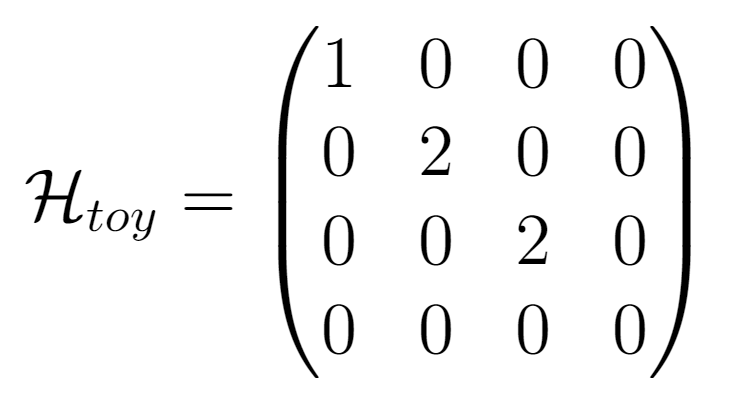}  \hspace{1cm}
(b)
\includegraphics[scale=0.3]{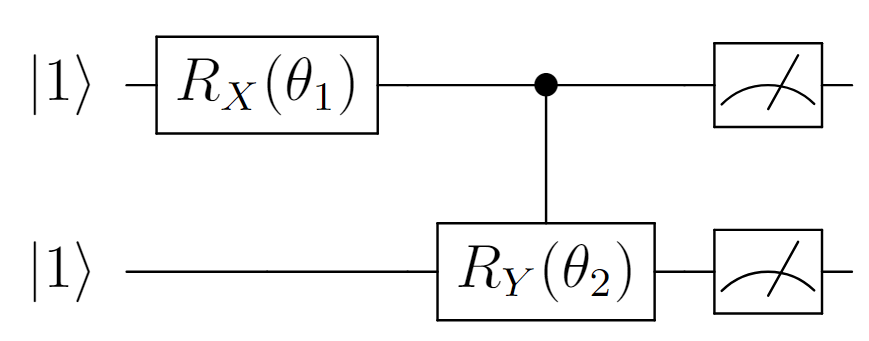} 
\newline
(c)
\includegraphics[scale=0.45]{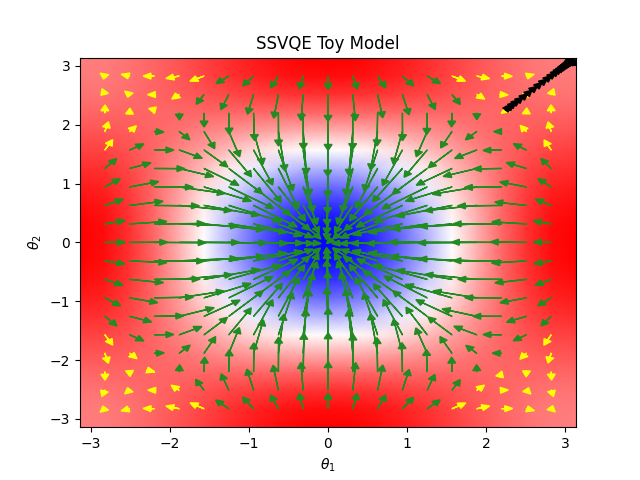} 
(d)
\includegraphics[scale=0.45]{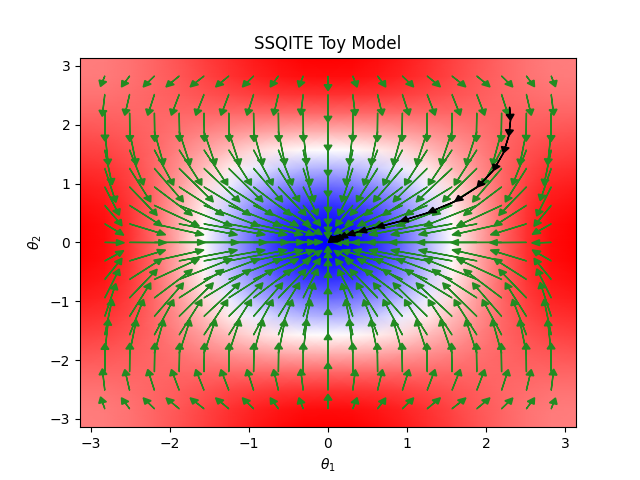}
\caption{Comparison of SSVQE and SSQITE on a toy model with local minima. (a) The 2-qubit toy Hamiltonian with local minima. (b) The two-parameter ansatz used. (c-d) SSVQE and SSQITE applied to the three lowest states of this Toy Hamiltonian and ansatz pair.  The orthogonal input states are $\ket{11}$, $\ket{00}$, and $\ket{01}$ respectively. The background coloring represents the weighted loss of the lowest three energy levels.  SSVQE is unable to escape the local minima labelled with yellow arrows, while SSQITE is easily able to escape this minima.  The black arrows represent a single run of SSVQE and SSQITE from the starting point $\theta_1 = \theta_2 = 2.3$, on the edge of the local minima.}
\label{fig:fig5}
\end{figure}

Lastly, the SSQITE algorithm was compared to SSVQE on the a simple toy Hamiltonian in Figure~\ref{fig:fig5}.  This comparison was done across the lowest three states in the toy two-qubit Hamiltonian, with ansatz shown in Figure~\ref{fig:fig5}b.  Using this toy model, it is shown that SSVQE can become trapped in local minima, whereas SSQITE is able to escape to the global minimum. In Figure~\ref{fig:fig5}, SSVQE becomes trapped in the local minimum located at $\theta_1 = \theta_2 = \pm\pi$, while SSQITE instead finds the global minimum at $\theta_1 = \theta_2 = 0$. As VarQITE applied to ground states has previously been demonstrated to have a resistance to local minima as compared to VQE with a gradient descent optimizer \cite{McArdle_2019}, this toy model demonstrates that this resistance holds even when VarQITE is extended to excited state algorithms such as SSQITE.

\section{Conclusions}
\label{sec:conclusions}
We have introduced the SSQITE method for computations of excited states using quantum devices. This method combines key aspects of the SSVQE and VarQITE methodologies. We demonstrated the capabilities of SSQITE by calculating the low-lying excited states of $\text{H}_2$ and $\text{LiH}$ molecules. The results showed robustness in avoiding local minima and excellent agreement with numerically exact calculations. We have also demonstrated the resistance of SSQITE to local minima through a simple toy model. Additionally, SSQITE is not sensitive to degenerate states, unlike folded-spectrum VQE or folded-spectrum VarQITE, which calculate excited states by altering the Hamiltonian to $(\mathcal{H} - E)^2$ \cite{Cadi_Tazi_2024, Tsuchimochi_2023}, where $E$ is the energy of interest.

We have shown that using VarQITE as a foundation for excited state algorithms offers potential benefits relative to VQE, since some local minima typically found during VQE gradient descent are absent in VarQITE \cite{McArdle_2019}. We have demonstrated that this advantage persists when applied to excited state algorithms.  Additionally, we anticipate that the subspace-search methodology implemented in SSQITE could also be applied to exploit the advantages in other algorithms such as the Quantum Iterative Power Algorithm (QIPA). QIPA uses an oracle which double-exponentiates the Hamiltonian $\alpha(-\tau\mathcal{H}) = e^{e^{-\tau\mathcal{H}}}$ in order to amplify the global minimum of any input state. This has been shown to require fewer iterations than VarQITE for quantum optimization of ground states \cite{Kyaw_2023}. This suggests that the combination of subspace-search and imaginary time quantum evolution methodologies could outperform other currently available algorithms for computations of excited states.

\section{Acknowledgments}
The authors acknowledge support from the National Science Foundation Engines Development Award: Advancing Quantum Technologies (CT) under Award Number 2302908. VSB also acknowledges partial support from the National Science Foundation Center for Quantum Dynamics on Modular Quantum Devices (CQD-MQD) under Award Number 2124511.

\section{Code availability}
The Python code for the SSQITE simulations is available at \href{https://colab.research.google.com/drive/1H1NAtboOIcNp2x2LTVIJCsRTz36Mkv7_#scrollTo=GV1V3p8MOU2V}{\textcolor{blue}{this link}}. %

\bibliography{biblio}

\end{document}